\documentclass[aps,twocolumn,showpacs,preprintnumbers,amsmath,amssymb]{revtex4}
\usepackage{graphicx}

\begin{document}

\title{Quantum logic gates using Stark shifted Raman transitions in a cavity}
\author{Asoka Biswas and G. S. Agarwal}
\affiliation{Physical Research Laboratory, Navrangpura, Ahmedabad-380 009, India}
\date{\today}

\begin{abstract}
We present a scheme to realize the basic  two-qubit logic gates such as quantum phase 
gate and controlled-NOT gate using a detuned 
optical cavity interacting with a three-level Raman system.
We discuss the role of Stark shifts which are as important as the terms
leading to two-photon transition. The operation of the 
proposed logic gates involves metastable states of the atom and hence is not affected by 
spontaneous emission. These ideas can be extended to produce multiparticle entanglement.
\end{abstract}

\pacs{03.67.-a, 03.67.Lx}

\maketitle

\section{Introduction}
The performance of a quantum computer relies on certain universal one-qubit and two-qubit 
logic gates. Any quantum computation \cite{qc} can be reduced 
to 
a sequence of these gates \cite{twoqubit,barenco}. There have been a number of 
experimental 
systems proposed as candidates for implementing these logic gates and many of these 
have been implemented. We may 
mention 
trapped ions \cite{ion}, cavity QED \cite{pelli,qed1,raus}, NMR \cite{nmr,cnot},
quantum dots \cite{cnot,qdot}, and neutral atoms in optical lattice \cite{atom}
as examples. Some of the basic two-qubit logic gates  are the conditional 
quantum phase gate (QPG) \cite{qpg}, controlled-NOT (CNOT) gate which is a 
universal two qubit gate \cite{barenco,cnot}, SWAP gate etc. It should be 
mentioned that a $\pi$ shift of QPG and appropriate rotation of the second 
qubit realize the CNOT gate. 

The QPG can be performed using a three-level atom interacting 
with a detuned cavity.
The two-photon transitions are especially attractive in this case as then one can work 
with long-lived ground states of the atom. In such a situation the excited state does not
participate in the transition and thus it is possible to minimize the effect of decoherence
associated with the finite lifetime of the excited state \cite{rempe, zoller_kimble}.
However, the two-photon transitions have complications associated with Stark shifts of the energy levels. 
The Stark shifts are quite natural to any two-photon process as one considers single-photon
transitions which are detuned from the intermediate levels. If one ignores Stark shifts, as is very
often done, then the nature of the two-photon process becomes similar to the one-photon process
and many of the results like Rabi oscillations carry over to two-photon processes. In this paper
we consider a situation where a three-level atom in $\Lambda$ configuration interacts with a 
bimodal cavity where the modes are highly detuned from the corresponding one-photon transition. 
We demonstrate the possibility of performing a number of logic operations (e.g., QPG, CNOT, SWAP) 
using two-photon Raman
transition. We show this in spite of the non-zero Stark shifts in the Raman transitions.

The structure of the paper is as follows. In Sec.~II, we present the model 
system and its theoretical description. In Sec.~III, we show how
different two-qubit logic gate operations can be performed   
using this model. In Sec.~IV we discuss the role of Stark shift
in quantum logic gate operations.

\begin{figure}
\scalebox{0.5}{\includegraphics{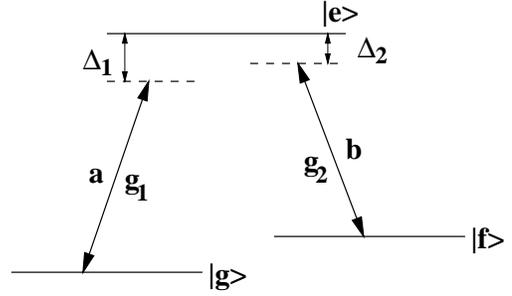}}
\caption{\label{fig1}Three-level atomic configuration with levels $|g\rangle$, 
$|f\rangle$, and $|e\rangle$ interacting with two orthogonal modes of the 
cavity, described by annihilation operators $a$ and $b$. Here $g_1$ and $g_2$ represent the 
atom-cavity coupling of the $a$ and $b$ modes with the corresponding 
transitions, $\Delta_i$'s ($i \in 1,2$) are the respective one-photon detunings.}
\end{figure}

\section{model}
Let us consider a three-level atomic configuration [Fig.\ \ref{fig1}]. The atom passes through a bimodal cavity. The modes with 
annihilation operators $a$ and $b$ interact with the 
$|e\rangle \leftrightarrow |g\rangle$ and $|e\rangle \leftrightarrow |f\rangle$ 
transitions, respectively.
 The Hamiltonian under rotating wave approximation can be 
written as
\begin{eqnarray}
H&=&\hbar\left[\omega_{eg}|e\rangle\langle e|+\omega_{fg}|f\rangle\langle f|+\omega_1a^\dag a+\omega_2b^\dag b\right.\nonumber\\
&&+\left.\left\{g_1|e\rangle\langle g|a+g_2|e\rangle\langle f|b+\textrm{h.c.}\right\}\right]
\label{fullhamil}
\end{eqnarray}

\noindent
where, $\omega_{lg}$ $(l\in e,f)$ is the atomic transition frequency, $\omega_i$
$(i\in 1,2)$ is the frequency of the cavity modes $a$ and $b$, and 
$g_i$ is the atom-cavity coupling constant. We assume $g_i$ to be real. 
The interaction Hamiltonian in 
the interaction picture can be written as 
\begin{equation}
H=\hbar \left[g_1|e\rangle\langle g|a e^{i\Delta_1 t}+g_2|e\rangle\langle f|be^{i\Delta_2 t}+\textrm{h.c.}\right],
\label{hamil1}
\end{equation}

\noindent
where, $\Delta_i=\omega_{eg,f}-\omega_i$ ($i\in 1,2$) is the  
one-photon detuning of the cavity modes. 
 
If the initial number of photons in the $a$ and $b$ modes are $n$ and $\mu$, 
respectively, then the state vector of the atom-cavity system can be expanded 
in terms of the possible basis states in the following way: 
\begin{equation}
|\psi(t)\rangle=c_1|g\rangle|n,\mu\rangle+c_2|e\rangle|n-1,\mu\rangle+c_3|f\rangle|n-1,\mu+1\rangle,  \label{wavefunc}
\end{equation}

\noindent
where, $c_i$'s $(i\in 1,2,3)$ are the probability amplitudes for the corresponding states. The amplitude equations can be obtained from the Schr{\"o}dinger 
equation as 
\begin{eqnarray}
\dot{d}_1&=&-ig_1\sqrt{n} d_2,\nonumber\\
\label{ampequ}\dot{d}_2&=&-i[g_1\sqrt{n}d_1+g_2\sqrt{\mu+1}d_3]-i\Delta_1 d_2,\\
\dot{d}_3&=&-ig_2\sqrt{\mu+1}d_2-i(\Delta_1-\Delta_2)d_3,\nonumber
\end{eqnarray}

\noindent
where the following transformations have been used:
\[c_1\rightarrow d_1,~~~~ c_2e^{-i\Delta_1 t} \rightarrow d_2,~~~~c_3 e^{-i(\Delta_1-\Delta_2)t}\rightarrow 
d_3.\]

We now work under the limit of large one-detunings. We assume that, $g_1=g_2=g$,
$\Delta_i \gg g$ and $(\Delta_1-\Delta_2)\ll g$. In this limit, we put $\dot{d}_2\approx 0$ and the amplitude 
equations (\ref{ampequ}) reduce to 
\begin{subequations}
\begin{eqnarray}
\dot{d}_1&=&\frac{ig^2\sqrt{n}}{\Delta_1}\left[\sqrt{n}d_1+\sqrt{\mu+1}d_3\right], \\
\dot{d}_3&=&-i(\Delta_1-\Delta_2)d_3+\frac{ig^2\sqrt{\mu+1}}{\Delta_1}\left[\sqrt{n}d_1+\sqrt{\mu+1}d_3\right]\;.\nonumber\\
\end{eqnarray}
\label{amp1}
\end{subequations}

\noindent
We note that the Eqs. (\ref{amp1}) can be obtained from an 
effective Hamiltonian given by
\begin{eqnarray}
H_{\textrm{eff}}&=&-\frac{\hbar g^2}{\Delta_1}\left[|g\rangle\langle g|a^\dag a+|f\rangle\langle f|b^\dag b\right]\nonumber\\
&&-\frac{\hbar g^2}{\Delta_1}\left[|g\rangle\langle f|a^\dag b+|f\rangle\langle g|a b^\dag\right]\nonumber\\
&&+\hbar(\Delta_1-\Delta_2)|f\rangle\langle f|.
\label{effham}
\end{eqnarray}

\noindent
Here the first two terms represent the Stark shifts and the next two terms
give the interaction leading to a transition from the initial state to the 
final state. The last term represents a shifting of the level $|f\rangle$ due
to the two-photon detuning. From the Hamiltonian (\ref{effham}), one can
easily see that the Stark shift terms are of the {\it same order of magnitude\/}
($=\hbar g^2/\Delta_1$) as the coupling term, 
and thus are as important as the coupling term and should be kept in further 
discussion. So one cannot ignore 
these Stark shift terms from the Hamiltonian.
The solution of Eqs.~(\ref{amp1}) is
\begin{subequations}
\begin{eqnarray}
d_1(t)&=&e^{i\nu t/2}\left\{\left[\cos{(\frac{\Omega t}{2})}+\frac{i}{\Omega}(\Delta_1-\Delta_2)\sin{(\frac{\Omega t}{2})}\right]d_1(0)\right.\nonumber\\
\label{sol1}&+&\left.\frac{2ig^2}{\Delta_1\Omega}\sin(\frac{\Omega t}{2})d_3(0)\right\};\\
d_3(t)&=&e^{i\nu t/2}\left\{\left[\cos{(\frac{\Omega t}{2})}-\frac{i}{\Omega}(\Delta_1-\Delta_2)\sin{(\frac{\Omega t}{2})}\right]d_3(0)\right.\nonumber\\
\label{sol3}&+&\left.\frac{2ig^2}{\Delta_1\Omega}\sin(\frac{\Omega t}{2})d_1(0)\right\};
\end{eqnarray}
\label{solns}
\end{subequations}

\noindent
where
\begin{eqnarray}
\Omega&=&\left[\left(\frac{2g^2}{\Delta_1}\right)^2+(\Delta_1-\Delta_2)^2\right]^{1/2},\nonumber\\
\nu&=&\left[\frac{2g^2}{\Delta_1}-(\Delta_1-\Delta_2)\right],
\end{eqnarray}
and we have considered $n=1$ and $\mu=0$. Under two-photon resonance 
condition $\Delta_1=\Delta_2=\Delta$, the solution reduces to 
\begin{subequations}
\begin{eqnarray}
d_1(t)&=&\frac{1}{2}[d_1(0)+d_3(0)](e^{i\theta}-1)+d_1(0)\;,\\
d_3(t)&=&\frac{1}{2}[d_1(0)+d_3(0)](e^{i\theta}-1)+d_3(0)\;,
\end{eqnarray}
\label{amp2}
\end{subequations}

\noindent
where $\theta=2g^2t/\Delta$.

We note that, if we exclude Stark shift terms from the Hamiltonian (\ref{effham}), 
and work under two-photon resonance (i.e., $\Delta_1=\Delta_2=\Delta$) then 
the effective Hamiltonian reduces to 
\begin{equation}
H'_\mathrm{eff}=-\frac{\hbar g^2}{\Delta}\left(S^-a^\dag b+S^+ab^\dag\right),
\label{effham1}
\end{equation}

\noindent
where $S^+=|f\rangle\langle g|$ and $S^-=|g\rangle\langle f|$ are the atomic 
two-photon creation and annihilation operators, 
respectively. The solution of the Schr\"odinger equations using this Hamiltonian is
\begin{subequations}
\begin{eqnarray}
d_1(t)&=&\cos\left(\frac{g^2t}{\Delta}\right)d_1(0)+i\sin\left(\frac{g^2t}{\Delta}\right)d_3(0)\;,\\
d_3(t)&=& \cos\left(\frac{g^2t}{\Delta}\right)d_3(0)+i\sin\left(\frac{g^2t}{\Delta}\right)d_1(0)\;,
\end{eqnarray}
\label{soln1}
\end{subequations}

\noindent
which represents a Rabi Oscillation of the vector $(d_1, ~d_3)$.

\section{Quantum logic gate operations}

In this section we will show how different kinds of two-qubit logic gates can 
be performed using the present model.

\subsection{QPG operation}

Let us first consider the solutions of Eqs.~(\ref{amp1}) under the 
total effective Hamiltonian (\ref{effham}) (with $\Delta_1\neq\Delta_2$),
 given by Eqs. (\ref{solns}). From these solutions  
 one can easily see that, if 
\begin{equation}
\frac{\Delta_1-\Delta_2}{g}=\frac{2}{(\Delta_1/g)}\;,
\end{equation}
then, for an interaction time $t=T$ defined by  
\begin{equation}
\label{qpgtime}gT=\frac{\pi}{\sqrt{2}}.\left(\frac{\Delta_1}{g}\right)=\frac{\sqrt{2}\pi}{(\Delta_1-\Delta_2)/g}\;,
\end{equation}
$d_1(t)$ becomes $-1$ for the initial condition $d_1(0)=1$.
At this particular interaction time, 
one can perform the following QPG operation:
\begin{eqnarray}
|0_a\rangle|0_b,g\rangle &\longrightarrow & |0_a\rangle|0_b,g\rangle,\nonumber\\
\label{qpg}|0_a\rangle|0_b,f\rangle &\longrightarrow & |0_a\rangle|0_b,f\rangle,\\
|0_a\rangle|1_b,g\rangle &\longrightarrow & |0_a\rangle|1_b,g\rangle,\nonumber\\
|0_a\rangle|1_b,f\rangle &\longrightarrow & -|0_a\rangle|1_b,f\rangle,\nonumber
\end{eqnarray}
which clearly involves the {\it atomic ground state\/} basis and the {\it Fock state\/} basis in $b$ mode.
It should be mentioned here that, we have verified the above analytical 
results by solving the Eqs.~(\ref{ampequ})
numerically for the amplitudes $d_i$'s also. The numerical results reveal that  
that the adiabatic approximation used in the present problem holds good.

Note that, in the above QPG operation we have considered $\Delta_1\neq\Delta_2$.
But if we work under the two-photon resonance condition ($\Delta_1=\Delta_2=\Delta$), 
then the solutions of Eqs. (\ref{amp1}) are given by Eqs. (\ref{amp2}).
From these solutions one can easily notice that the time dependent amplitude of the initial 
state 
$|0_a\rangle|f,1_b\rangle$ is now $(e^{2ig^2T/\Delta}+1)/2$ 
[see Eq.~(\ref{amp2}b)], which never reaches (-1). Rather for a 
certain choice of $2g^2T/\Delta$($=\pi/2$), this becomes $e^{i\pi/4}/\sqrt{2}$. 
This clearly shows that, the system no longer remains in that state (as 
obvious from the factor $1/\sqrt{2}$). Transition takes place to another basis 
state $|1_a\rangle|g,0_b\rangle$. In this way, working with the total effective
Hamiltonian under two-photon resonance, 
one cannot perform phase gate operation. Thus 
in the present model, QPG can be performed 
successfully {\it only by avoiding the two-photon resonance condition\/}. 

We emphasize that the QPG operation discussed above 
involves the cavity mode $b$ as well as the ground metastable states
($|g\rangle$ and $|f\rangle$) of the atom, transition between 
which is dipole-forbidden. This is unlike the
case in \cite{raus}, where the authors used two Rydberg states (states with very large
quantum numbers) which are dipole-coupled. Thus, the QPG operation, discussed in the present paper
is not affected by any kind of decoherence due to spontaneous emission of the 
atomic levels, though it is limited by the cavity life-time as all operations like
the storage of the photons after initial preparation and detection of the photonic qubit are to be done within the cavity life-time. To realize the QPG against the
cavity decay, one must has to meet the condition $\pi\Delta_1\kappa/\sqrt{2}g^2<1$, which directly follows from the condition $T<\kappa^{-1}$ and Eq.~(\ref{qpgtime}), where $\kappa$ is the cavity decay constant. A possible parameter zone 
to satisfy the aove condition is $\Delta_1=10g$ and $\kappa=0.01g$. This, though challenging for an optical cavity, can be expected to reach very soon.

Note that if we consider a third atomic metastable state $|k\rangle$ which
is an auxiliary state, it
is possible to perform the following QPG involving the {\it atomic metastable 
states\/} ($|g\rangle$ and $|k\rangle$) and the two-mode photonic basis 
($|g_R\rangle\equiv |0_a,1_b\rangle$ and $|e_R\rangle\equiv |1_a,0_b\rangle$);
\begin{eqnarray}
|k\rangle|g_R\rangle&\longrightarrow &|k\rangle|g_R\rangle\;,\nonumber\\
\label{qpg_alt}|k\rangle|e_R\rangle&\longrightarrow &|k\rangle|e_R\rangle\;,\\
|g\rangle|g_R\rangle&\longrightarrow &|g\rangle|g_R\rangle\;,\nonumber\\
|g\rangle|e_R\rangle&\longrightarrow &-|g\rangle|e_R\rangle\;,\nonumber
\end{eqnarray}
at the time defined by (\ref{qpgtime}).
We should mention here that there are several other schemes, which use 
non-interacting levels to perform logic gates in two-atom basis.
Our model is quite different from other schemes \cite{pachos,plenio,yi}.
Note that we use a single 
atom interacting with a two-mode cavity unlike the cases cited above, which 
use two atoms interacting with a single-mode cavity. 

\subsection{CNOT operation}
A CNOT gate can be implemented from a QPG, through a rotation of the second
qubit before and after the QPG operation. We choose the atom as the second
qubit in the present problem. By applying the Hadamard transformation on the
atomic state, before and after the QPG operation (\ref{qpg}), we obtain the following
CNOT operation:
\begin{eqnarray}
|0_a\rangle|0_b,g\rangle&\stackrel{\hat{C}}{\longrightarrow}&|0_a\rangle|0_b,g\rangle \;,\nonumber\\
|0_a\rangle|0_b,f\rangle&\stackrel{\hat{C}}{\longrightarrow}&|0_a\rangle|0_b,f\rangle \;,\\
|0_a\rangle|1_b,g\rangle&\stackrel{\hat{C}}{\longrightarrow}&|0_a\rangle|1_b,f\rangle \;,\nonumber\\
|0_a\rangle|1_b,f\rangle&\stackrel{\hat{C}}{\longrightarrow}&|0_a\rangle|1_b,g\rangle \;,\nonumber
\end{eqnarray}

\noindent
where $\hat{C}$ represents the CNOT operation here. Here the Hadamard transformation
on the atomic qubit  states $|g\rangle$ and $|f\rangle$
can be implemented by applying two resonant cw fields
with equal Rabi frequencies in the respective transitions of the atom. We
identify the field qubit as controlling qubit and the atomic qubit as the controlled qubit.
We note that recently,
DeMarco {\it et al.\/} have demonstrated a CNOT gate operation in a single
trapped ion interacting with a single Raman pulse \cite{wineland}.

\subsection{\label{sec:swap}SWAP gate operation}

In order to arrive at SWAP gate, 
we rewrite the Hamiltonian (\ref{effham}) as an interaction between the two 
``qubits" in the following way:
\begin{eqnarray}
\tilde{H}_{\textrm{eff}}&=&-\frac{\hbar g^2}{\Delta_1}\left[S^+R^-+S^-R^+-2S^zR^z+\frac{1}{2}\right]\nonumber\\
&+&\hbar(\Delta_1-\Delta_2)\left(S_z+\frac{1}{2}\right)
\label{neweff}
\end{eqnarray}

\noindent
where, 
\begin{eqnarray}
&&S^+=|f\rangle\langle g|\;,\;S^-=|g\rangle\langle f|\;,\;S^z=\frac{1}{2}(|f\rangle\langle f|-|g\rangle\langle g|)\;,\nonumber\\
&&R^+=a^\dag b\;,\;R^-=ab^\dag \;,\;R^z=\frac{1}{2}(a^\dag a-b^\dag b)\;.
\end{eqnarray}
Here we identify a single photon in a two-mode cavity as an
effective qubit with the two possible states $|e_R\rangle$ and
$|g_R\rangle$.
The field operator $R$ acts like a spin-1/2
operator in this basis. 
We now assume that two-photon resonance condition
is satisfied ($\Delta_1=\Delta_2=\Delta$) so that the effective Hamiltonian reads as  
\begin{equation}
\tilde{H}_{\textrm{eff}}=-\frac{2\hbar g^2}{\Delta}[S^xR^x+S^yR^y-S^zR^z+\frac{1}{4}]\;,
\label{ham200}
\end{equation}

\noindent
where $S^x=(S^++S^-)/2$ and $S^y=(S^+-S^-)/2i$.
This signifies a spin-exchange interaction between two spin-1/2 particles. This kind
of interaction is always responsible for swap operation. 
Defining the unitary operation as
\begin{eqnarray}
U&=&\exp(-i\tilde{H}_{\textrm{eff}}t/\hbar)\nonumber\\
&=&\exp[i\theta (S^xR^x+S^yR^y-S^zR^z+\frac{1}{4})]\;,
\label{unit}
\end{eqnarray}

\noindent
where, $\theta=2g^2t/\Delta$, a SWAP gate can be performed 
for $\theta=\pi$ [$U_{\textrm{SW}}=U(\theta=\pi)$]:
\begin{eqnarray}
|g\rangle|g_R\rangle&\stackrel{U_{\textrm{SW}}}{\longrightarrow}&|g\rangle|g_R\rangle\;,\nonumber\\
|g\rangle|e_R\rangle&\stackrel{U_{\textrm{SW}}}{\longrightarrow}&-|f\rangle|g_R\rangle\;,\\
|f\rangle|g_R\rangle&\stackrel{U_{\textrm{SW}}}{\longrightarrow}&-|g\rangle|e_R\rangle\;,\nonumber\\
|f\rangle|e_R\rangle&\stackrel{U_{\textrm{SW}}}{\longrightarrow}&|f\rangle|e_R\rangle\;.\nonumber
\end{eqnarray}

\subsection{Role of phases of the coupling constants in SWAP gate}

In all the above calculations, we have assumed that the field coupling constants
$g_1$ and $g_2$ are equal and {\it in phase\/}. However if they are not so, the general 
expression for the Hamiltonian [Eq.~(\ref{effham})] under two-photon resonance 
would be
\begin{eqnarray}
H'&=&-\frac{\hbar}{\Delta}\left[|g_1|^2|g\rangle\langle g|a^\dag a+|g_2|^2|f\rangle\langle f|b^\dag b\right.\nonumber\\
&+&\left.g_1g_2^*|g\rangle\langle f|a^\dag b +g_1^*g_2|f\rangle\langle g|ab^\dag\right]\;.
\end{eqnarray}
Now if we impose the conditions 
\begin{equation}
|g_1|=|g_2|=g,\;\;g_2=-g_1\;,
\end{equation}
then the above Hamiltonian can be written as 
\begin{eqnarray}
H'&\equiv& \frac{2\hbar g^2}{\Delta}[S^xR^x+S^yR^y+S^zR^z-\frac{1}{4}]\nonumber\\
&=&\frac{2\hbar g^2}{\Delta} [\vec{S}.\vec{R}-\frac{1}{4}]
\end{eqnarray}
instead of Eq.~(\ref{ham200}). Then the corresponding unitary operation $U'$ 
becomes
\begin{eqnarray}
U'&=&e^{-iH't/\hbar}=\exp{\left[-i\theta \left(\vec{S}.\vec{R}-\frac{1}{4}\right)\right]}\nonumber\\
&=&[1+(e^{-i\theta}-1)\hat{P}]e^{i\theta}\;.
\end{eqnarray}
Here $\hat{P}=3/4 +\vec{S}.\vec{R}$ is the projection operator with the eigenvalues ) and 1.  
The SWAP gate of Sec.~\ref{sec:swap} can also be implemented using the above 
unitary operator for $\theta=\pi$.
It is also very interesting to note that not only for a particular phase relation between $g_1$ and $g_2$, but for any arbitrary phase between them, the SWAP
gate works in the following way:
\begin{eqnarray}
|g\rangle|g_R\rangle&\longrightarrow&|g\rangle|g_R\rangle\;,\nonumber\\
|g\rangle|e_R\rangle&\longrightarrow&-e^{i\phi}|f\rangle|g_R\rangle\;,\\
|f\rangle|g_R\rangle&\longrightarrow&-e^{-i\phi}|g\rangle|e_R\rangle\;,\nonumber\\
|f\rangle|e_R\rangle&\longrightarrow&|f\rangle|e_R\rangle\;,\nonumber
\end{eqnarray}
where $\phi$ is defined through the relation $e^{i\phi}=g_1g_2^*/|g_1|.|g_2|$.

We should emphasize that all these universal logic gates are the key resource 
in quantum computation.  
Our method and system can also be used to prepare two-particle and three-particle 
entangled states involving 
metastable states of the atoms. This can be done by sequentially addressing the atoms by the
two-mode cavity under two-photon resonance condition (cf. \cite{2entangle,raus00}). 

\section{role of Stark shifts in quantum logic operations}

Next we investigate the role of the Stark shift term in performing 
the logic gates. Let us consider the Hamiltonian (\ref{effham1})
under two-photon resonance condition which {\it excludes the Stark shift term}. 
From the corresponding solutions of (\ref{soln1}) for the probability amplitudes,
one can obtain the following QPG operation with a $2\pi$-pulse ($2g^2T/\Delta=2\pi$):
\begin{eqnarray}
|0_a\rangle|0_b,g\rangle &\longrightarrow & |0_a\rangle|0_b,g\rangle,\nonumber\\
|0_a\rangle|0_b,f\rangle &\longrightarrow & |0_a\rangle|0_b,f\rangle,\\
|0_a\rangle|1_b,g\rangle &\longrightarrow & |0_a\rangle|1_b,g\rangle,\nonumber\\
|0_a\rangle|1_b,f\rangle &\longrightarrow & -|0_a\rangle|1_b,f\rangle.\nonumber
\end{eqnarray}
But as soon as we keep the Stark shift term into the Hamiltonian 
[see Eq.~(\ref{effham})] and continue to work under the two-photon 
resonance condition, one cannot achieve phase gate operation.  
It should be borne in mind that one cannot ignore the Stark shift term 
as they are as important as the coupling term in the Hamiltonian.
Our Sec. IIIA shows how to perform the QPG in spite of the Stark shift term.  
We used the extra freedom provided by two-photon detunings in our model.
We should mention here that many authors have 
utilized an additional field to cancel the unwanted Stark shifts 
\cite{stark1,stark2,stark3}.
We also note that in the context of other models, Stark shifts have been used for 
two-qubit logic \cite{ion_gate}, the Deutsch-Jozsa algorithm \cite{chandan},
and quantum holography \cite{holo_gsa}.

Note that, recently, Solano {\it et al.\/} have reported a QPG operation based on the interaction 
of a three-level atom in ladder configuration and two modes of a cavity (each 
mode can have either zero or one photon). The cavity modes are highly detuned 
from single photon transition (see Fig. 2, \cite{solano}), but are 
two-photon resonant. 
They have shown how excluding the self-energy terms in the effective Hamiltonian,
one can perform a QPG operation in photonic Hilbert space
 ($ |0,0\rangle \rightarrow |0,0\rangle, |0,1\rangle \rightarrow |0,1\rangle, |1,0\rangle \rightarrow
|1,0\rangle, |1,1\rangle \rightarrow -|1,1\rangle$).

\section{conclusions}
In conclusion, we have presented a system where a three-level atom in 
$\Lambda$-configuration interacts with a high-Q bimodal optical cavity, 
with the cavity 
modes being highly detuned from the corresponding single-photon transitions. 
We have shown how a variety of logic operations can be performed using the 
ground states of the atoms. The decoherence associated due to spontaneous emission
is thus negligible, though the quality of the cavity would lead some decoherence.
We further emphasize that Stark shifts are systematically included in our case.
Further the present system can be used to prepare  
bipartite and tripartite entangled states involving the metastable states of the atoms.

\end{document}